\documentstyle[preprint,psfig,aps]{revtex}
\begin{document}
\draft
\title{Application of the time-dependent local-density 
approximation to conjugated molecules}
\author{K. Yabana\footnote{E-mail address yabana@nt.sc.niigata-u.ac.jp}}
\address{Graduate School of Science and Technology, Niigata University\\
Niigata 950-21, Japan\\
and\\}
\author{G.F. Bertsch\footnote{E-mail bertsch@phys.washington.edu}}
\address{Physics Department and Institute for Nuclear
Theory\\
University of Washington, Seattle, WA 98195 USA
}
\date{text/conjug/conjug.tex; June 11, 1998}
\maketitle
\def\C60{C$_{60}$}
\def\eq#1{eq. (\ref{#1})}
\def\pipi{$\pi-\pi^*$}
\def\sigsig{$\sigma -\sigma^*$}
\begin{abstract}
The time-dependent local-density approximation 
(TDLDA) is applied to the optical response of 
conjugated carbon molecules in the
energy range of 0-30 eV, with calculations given for
carbon chains, polyenes, retinal, benzene and \C60.
The major feature of the spectra, the collective
\pipi~transition, is seen at energies ranging
from below 2 to 7 eV,
and is reproduced by the theory to a few tenths of an eV
with a good account of systematic trends.  However, there
is some indication that TDLDA predicts too much fragmentation
of the strength function in large molecules.
Transition 
strengths are reproduced with a typical accuracy of 20\%.
The theory also
predicts a broad absorption peak in the the range 15-25 eV,
and this feature agrees with experiment in the one case
where quantitative data is available (benzene). 
\end{abstract}
\def\be{\begin{equation}}
\def\ee{\end{equation}} 

\section{Introduction}

Mean field theory is widely used in chemistry and physics, giving
an approximation that is very robust in its domain of validity.
Particularly successful is the density-functional theory\cite{jo89},
which treats the electron-electron interaction in a local-density
approximation (LDA), i.e. by adding a density-dependent contact
term to the Hartree Hamiltonian.  This is similar to the
$\alpha X$ approximation, but the contact term
includes an approximate treatment of correlation effects as well
as the exchange interaction.
The theory is commonly used to calculate ground state structures of
molecules and condensed systems.  A corresponding theory for 
excitations is the time-dependent local-density approximation (TDLDA),
which has also been applied to a number of systems
\cite{ru96,ya94,ko93,lu94,lu95,fe92,ya96,ya97,ja96,bl95,ru96,za80}.
In
this work, we will attempt to make a systematic survey of the predictions
of the TDLDA for conjugated carbon molecules.  Our calculations will
be {\it ab initio} from the point of view of the electrons, since we
take the electron Hamiltonian from well-known prescriptions used in 
{\it ab initio} density-functional Hamiltonians.  However, we will not
attempt to use the Hamiltonian to arrive at the nuclear coordinates;
the ground state problem has been well studied by others and our
interest here is in the excitations.  In the next section we review
the TDLDA theory and our numerical implementation of it.  Then we
discuss the predictions and comparison with experiment for carbon
chains, polyenes, the aromatic ring, and \C60.

\section{Theory}

\subsection{Formal aspects}

The TDLDA theory is closely related to the time-dependent 
Hartree-Fock equation introduced by Dirac\cite{di30}.  That theory
can be derived from a variational principle,
\be
\delta \int\,dt\, \langle\Psi|H-i\hbar{\partial\over\partial t}|\Psi
\rangle=0.
\ee
Here $\Psi$ is restricted to be a Slater determinant of 
single-electron wave functions $\phi_i$, $\Psi={\cal A} \prod_i \phi_i$ .
The corresponding static variational principle is the usual energy
minimization principle from which one derives the Hartree-Fock equations.
The LDA theory is obtained in the same way, using the LDA energy
functional instead of $\langle\Psi|H|\Psi\rangle$ in eq. (1).  The derived
TDLDA equations of motion are given by
\be
\label{tdlda}
-{\hbar^2 \nabla^2\over 2 m}\phi_i + V_{ion} \phi_i +e^2\int\,dr'
{n(r')\over |r-r|}\phi_i + V_{xc}[n(r)]\phi_i = i\hbar{\partial\over\partial t}
\phi_i
\ee
where $n(r)=\sum_i |\phi_i(r)|^2$ is the electron density.  The $V_{xc}$
is the exchange-correlation potential, related to the exchange-correlation
energy $v_{xc}$ by $V_{ex}=d v_{xc}/d n$.

Several general remarks can be made here.  Practitioners of the
density-functional theory often claim justification for
the Kohn-Sham equations based on formal existence theorems.
Theorems can also be proven for the time-dependent case\cite{pe96},
but our own partiality to the theory is based on more pragmatic considerations.
First, if the time dependence is slow, the TDLDA should be as good as
the static theory because the motion would be adiabatic and governed
by the same Hamiltonian.  Second, integrals of the response over 
frequency obey sum rules, and these are satisfied in principle by
the small-amplitude TDLDA.
Since the sum rules are most sensitive to the high
frequency behavior, we see that the TDLDA is good at
both extremes, and therefore promising for describing the middle
ground.

The next remark concerns the domain of applicability of the theory.
The TDLDA describes the system by a single Slater determinant
which can be expressed in a particle-hole basis as a particle-hole
operator on the ground state.  The theory thus does not contain
the degrees of freedom necessary to describe
excitations that have a more complex character than single-electron
excitations.  Not only will the experimental spectrum be 
more complex due to mixing of states of different particle-hole character
but the coupling to vibrations will broaden the transitions on an
energy resolution scale of a few tenths of an eV.  
However, even though the TDLDA only includes single-electron 
degrees of freedom,
it is far superior to the static single-electron theory because
it includes dynamic screening effects.  The dynamic screening
arises directly from the time dependence of the mean field potential.
It shifts transition strength up out of the lowest particle-hole excitations
into a higher frequency range, of course preserving the sum rule.
This can be a very large effect; in the case of \C60 considered
below the strength of the lowest excitations are reduced by 
an order of magnitude.  In general, what we expect from the
TDLDA is the frequency location and strengths of the strong
transitions, but not details of its fragmentation into the
eigenstates of the many-electron Hamiltonian.

There are several implementations of the
TDLDA, particularly in the small-amplitude limit, which is all that
concern us unless nonlinear optical properties are wanted. 
The real-time method has been applied in other fields to describe
the small amplitude response\cite{de92,vr97}.
We emphasize that the time-dependent method we use here is
equivalent to the response function method used in ref. \cite{bl95,ja96,ru96}
and the matrix RPA method used in ref. \cite{fe92,ko93,ya93}, provided only
that the amplitude of the motion is small.  Thus in principle
the results depend only on the assumed energy density function,
and not on the particular method used to solved the equations.
Different methods are advantageous under different circumstances.
The matrix RPA is the only one of these methods that can
treat nonlocal interactions such as the Fock term.  However, the 
matrix RPA
method uses a particle-hole representation of the perturbed 
wave function, which becomes inefficient for systems having
a large number of particles.  The
arguments comparing the numerical efficiency of different methods 
is given in ref.~\cite{ya97}.
In brief, the basic object to be computed in the
response function method is a $D\times D$ matrix, where $D$
is the dimensionality of the basis.  In the time-dependent methods, 
the numerical object is the set of $N$ vectors of dimensionality
$D$, where $N$ is the number of electrons.  The computational
effort scales with the size of the system as $ND$ for the
time-dependent method with a sparse Hamiltonian, compared with
$D^3$ for the inversion of the matrix in the response function
methods.

\subsection{Numerical aspects}

The important decisions in a numerical implementation of the TDLDA
are the choice of basis, the choice of a local density function,
and the choice of the basic quantity calculated: time-dependent
wave function, energy eigenstates, or frequency response.  We consider
these aspects in turn.  

\subsubsection{Basis}
The three most popular representations for the electron wave function
are localized functions such as Gaussians or atomic orbitals
(e.g., ref.~\cite{fe92}), planes 
waves, and more recently, spatial meshes\cite{ch94}.  
We solve the TDLDA equations using a uniform spatial mesh, following
closely the technique introduced in ref. \cite{fl78}.
An important advantage of spatial meshes is that the single-electron
Hamiltonian is a sparse matrix.  This allows one to consider vectors
for the particle wave functions with many tens of thousands of points.  Of
course, a uniform spatial mesh has the disadvantage that the mesh
size is determined by requirements of small regions next to the ion
centers.  This is alleviated somewhat by the use of pseudopotentials
(see below).  The important numerical
parameters associated with the spatial mesh are the mesh spacing 
$\Delta x$ and the total number of mesh points $M$.  Thus the molecules are calculated
in a volume $V\sim M (\Delta x)^3$.  The boundary condition at the
surface of the volume is determined when the numerical representation
of the differential operator in the Hamiltonian is constructed.  For 
most of our calculations, the operator takes the wave function to be
zero on the mesh points outside the volume.  We use a higher-order
difference operator in our program, as was done in ref. \cite{fl78,ch94}.  
The required mesh spacing depends on atoms involved; with carbon
we use $\Delta x=0.3$~\AA~which gives a HOMO-LUMO gap to within
0.1 eV of the converged value for the Kohn-Sham equations.  The geometry
of the volume represented by the mesh is determined by requiring the
edge to be at least 4~\AA~from any ionic center.  The number of
points needed ranges from 19000 in C$_2$H$_4$ to 103000 in C$_{28}$H$_{30}$.

\subsubsection{Energy density function}
For our energy density function, we use the widely applied
exchange-correlation energy given by ref.~\cite{ce80,pe81}.  One can introduce
more complicated energy functions that have gradient corrections
or self-interaction corrections, but it was not clear that there
would be an advantage in using a more complicated form.

In addition, we replace the ionic potential by a pseudopotential
to eliminate the core electrons from the theory. 
In principle the theory should be used with all electrons, but
in structure calculations the core electrons are passive and
it is a computational waste to treat them explicitly.  We
use a standard prescription for constructing the pseudopotentials
of the ions\cite{tr91}.  This procedure gives a pseudopotential
that is nonlocal, depending on the orbital angular momentum of the
electron.  This has consequences for the 
sum rules\cite{al97}, but we believe these are well under control\cite{ya98}.
Numerically, we implement the nonlocality using a separable 
approximation as described by ref.\cite{kl82}.

\subsubsection{Ground state}
The first part of the calculation is to construct the electronic
ground state of the molecule with the same LDA Hamiltonian that will
be used for the time-dependent calculation.  We use an iterative
conjugate-gradient method for this part of the calculation.  Since
the computation time taken just a small fraction of the 
total ($\sim 10\%$), it was not worthwhile to optimize this step\cite{sa96}.  It
is very useful to compute at the same time the energies of occupied
and unoccupied orbitals near the Fermi surface.  Unlike the
Hartree-Fock energies, the LDA single-particle energy differences
provide a good approximation to particle-hole excitation energies
in the molecules.  The independent-particle model of the
excitation spectrum is based on these energies and the matrix
elements of the independent particle-hole states.

\subsubsection{Time-dependent equations}
Once one has the wave functions for the occupied ground state orbitals
$\phi^{(0)}_i$, they are perturbed by multiplying them by a phase factor,
\be
\phi_i(0)= e^{ikx}\phi^{(0)}_i.
\ee
Physically, this gives the effect of a short-duration electric 
field acting on the electrons.  
The wave functions are then evolved by
the time-dependent equation of motion, eq. (\ref{tdlda}).  Physical
observables are extracted from the time-dependent wave function
by Fourier transformation.  A basic quantity to describe the molecule's
interactions is the dynamic polarizability
$\alpha(\omega)$.  This is essentially given by the
Fourier transform of the 
time-dependent dipole moment:
\be
\alpha(\omega) = {e^2\over \hbar }
\int \,dt\, e^{-i\omega t}\sum_i\langle\phi_i(t)|x|
\phi_i(t)\rangle.
\ee
Another useful quantity is the strength function $S(\omega)$, whose integral
is the total oscillator strength. This is defined
\be
S(\omega) = {2 m\omega\over \pi\hbar^2 k}
 \int\,dt\, \sin \omega t \sum_i \langle\phi_i(t)|x|
\phi_i(t)\rangle.
\ee
We shall use the symbol $n_e$ (effective number of electrons)
for integral of $S$ when we consider
the excitation with a specific orientation of the electric field
with respect to the molecule.  The usual oscillator strength $f$
is the same quantity averaged over molecular orientations.

The numerical integration requires as a parameter the 
time step  $\Delta T$ over which the single-particle Hamiltonian
is treated as static.  Within the time step the equations are 
solved by a predictor-corrector method
as described in ref.~\cite{fl78}.  The required size of the time step
depends on the energy scale of the Hamiltonian.  Thus one expects
a dependence
$\Delta T \sim (\Delta x)^2$.  As a practical criterion,
we determine $\Delta T$ by requiring the norm of the wave function to be
accurately conserved.  When that is the case, the energy is also conserved to good
accuracy.  We found that $\Delta T=0.001 \hbar$/eV is needed for carbon structures
with a mesh size of $\Delta x = 0.3$~\AA.

Another numerical parameter is the length of the time integration $T$.
The effective resolution of the strength function is determined by 
this quantity.  In the timing tests we show below, the integration time
is $T=10~\hbar$/eV, corresponding to a resolution
of $\hbar/T=0.1$ eV.  In the graphs of the strength functions shown later
we
have integrated to $T=30-40~\hbar$/eV, giving peaks having widths of less
than 0.1 eV.
Since the accuracy of the TDLDA does not
approach this value, there is no benefit to integrate 
to longer times.  However, one must remember that the displayed strength 
functions have a spurious width due to the finite integration time.  
In principle, the TDLDA strength function is infinitely 
sharp below the
ionization threshold.

As mentioned earlier, an advantage of the real-time implementation of the
TDLDA is that the computing
effort scales with the number of particles $N$ and the spatial 
dimension $D$ as $ND\sim N^2$.  Methods
using a particle-hole representation or the response scale
as a higher power, at least if one applies the methods naively without
truncation of matrix diagonalization or inversion operations.  
We have examined the scaling of our computation time
for the polyenes, which discussed in detail in Sec.~4 below.  The crosses 
in Fig.~1 show the computation
time to set up the ground state wave function for series ethylene, butadiene,
...,C$_{28}$H$_{30}$, using 1500 iterations of the Kohn-Sham 
equations.  The abscissa gives the number of valence electrons
in the calculation.  The dashed line shows the dependence according to
$t\sim N^2$ scaling.  One sees that the $N^2$ scaling applies quite
nicely to the larger sized molecules\footnote{Our 
algorithm to construct the ground state requires repeated orthogonalization
of the single-particle orbitals, an operation that scales as 
$N^3$.  Evidently, the prefactor
for this operation is not large enough to make it noticeable.}.
The computation time for solving the time-dependent
equation starting from the ground state is shown by the open triangles in
Fig.~1.  Here we
used 10,000 time steps of 0.001 $\hbar$/eV to obtain the real-time
response over an interval of 10 $\hbar$/eV.  From the figure, we see that
the scaling is even a little weaker than $N^2$.  In fact the number of
points $D$ needed to represent the wave function scales more slowly than
$M\sim N$ due to surface effects.

\section{Carbon chains}

   The first carbon structures we applied the theory to are linear 
carbon chains.  The results were already reported in \cite{ya97},
but for completeness we summarize them here.  In that study we
assumed that the chains were straight with evenly spaced carbon
atoms.  The interesting quantities are the HOMO-LUMO gap 
\be
\Delta e
= e_{LUMO}-e_{HOMO} 
\ee
 and
the behavior of the collective $\pi-\pi^*$ excitation.  The 
HOMO-LUMO gap for molecules in the range C$_3$-C$_{20}$ is well
described by the simple parameterization
\be
\label{1d}
 \Delta e   = {A \over N+1}
\ee
with $A=17.5$~eV for even chains and $A=16$~eV for odd chains.  
This corresponds to the H\"uckel model with a hopping parameter 
$\beta=2.8$~eV.  Experimentally, transitions have been observed
at these energies for even-$N$ chains with $6\le N\le 12$.

The TDLDA predicted that the collective excitation was a single, sharp
state with a energy that varies more slowly with $N$ than eq.~(\ref{1d}).
A better fit can be made with the following functional form\cite{mi59}
\be
\label{lnN}
E =   A { \sqrt{ln N}\over N}.
\ee
The large-$N$ fit to eq.~(\ref{lnN}) has $A=31.7$~eV with no difference
between odd and even $N$.  Configuration-interaction quantum
chemistry calculations have been done for $N=3,5$ and 7\cite{pa88,ko95},
and the energies of the predicted collective excitations agree well with
TDLDA.   Experimentally, transitions have been observed 
for odd-$N$ molecules with $N=3,7,9,11,13$ and  15.   The systematics
follows eq.~(\ref{lnN}) with a coefficient that is 10\% lower, $C\approx28.3
$.  

The oscillator strength associated
with the $\pi-\pi^*$ transition was found to be the same in the independent
electron calculation as in the TDLDA.  This implies that the screening
due to the $\sigma$-electrons is weak.  The longitudinal oscillator 
strength was found to depend on the number of atoms in the chain roughly as
\be
\label{f-chain}
n_e = C (N-1)
\ee
with $C\approx 1.6$.  This oscillator strength has a physical interpretation
as the number of electrons in the $\pi$ manifold of states.  For odd-$N$
chains, that number is $2N-2$, about 20\% larger than eq.~(\ref{f-chain}).

\section{Polyenes}

  We next consider the polyenes with alternating double
and single bonds which have been studied in empirical
models\cite{sc76} and well as with the linear response\cite{lu94,lu95}.
Unlike the
small carbon chains, the polyene bonds lengths are not equal.
As a consequence the 
HOMO-LUMO gap goes to a finite value
rather than to zero as in eq.~(\ref{1d}).  Our structures 
have bond lengths of 1.45~\AA~for C-C bonds and 1.34~\AA~
for double C=C bonds, as is commonly assumed\footnote{However, ref. \cite{lu94}
quote a double C-C bond length of 1.544~\AA.}.  The
C=C-C and C=C-H bond angles are taken to be 124$^\circ$ and 119$^\circ$
respectively, in the all-{\it trans} conformation.
\def\c8h10{C$_8$H$_{10}$}
We first discuss a typical case, \c8h10.  Fig.~2 shows on left the
energies of the $\pi$ orbitals in the LDA calculation of the
ground state.  
On the right-hand side is
the spectrum of the H\"uckel model which will be discussed below.
Both the HOMO and the LUMO are in the $\pi$ manifolds indicated
in the Figure;
the HOMO-LUMO energy gap is 2.58 eV in the LDA.  The HOMO-LUMO
transition has an oscillator strength of $n_e=5.3$ in the LDA.
For comparison, the strength would be $n_e=8$ if that transition
absorbed all of the strength of the $\pi$ manifold, since there
are 8 electrons in the occupied $\pi$ orbitals.  The 35\% reduction from 
the nominal value is more severe than in the carbon chains; evidently 
the longer bond distances
in the polyenes cause some fragmentation of the strength.  

Next we show in Fig.~3 the TDLDA response of \c8h10 from zero
energy to 30 eV excitation.  One sees the
strong $\pi-\pi^*$ transition at 3.86 eV, with a strength
$n_e=3.9$.  The slight reduction of strength compared to the
single-electron picture is due to the screening effect of the
$\sigma$ electrons.  In Fig.~3 one also sees the
\sigsig~ transitions as the broad distribution between
10 and 20 eV.  The strength integrated up to 30 eV is
$n_e=30$; integrating up to 100 eV gives a strength of 41 which
compares well to the number of valence electron in \c8h10,
namely 42.  

We next turn to the systematics as a function of the length of 
the polyene molecule.  Fig.~4 shows the HOMO-LUMO gap and the
excitation energy of the strong $\pi-\pi^*$ transition as a
function of the number of carbon atoms. The HOMO-LUMO gap
systematics is compared to two functional forms in the figure.
The dot-dash line shows an $N^{-1}$ dependence as in eq.~(7), which
is clearly wrong. The dashed line is a fit to the generalized
H\"uckel model with two hopping parameters $\beta_s$ and $\beta_d$
corresponding to
the single and double CC bonds, respectively.  For the fit, 
the parameter values are 
\be
  \beta_s=-2.27 {\rm~eV~and}~~ \beta_d=-2.80~ {\rm eV.}
\ee  
These are somewhat larger than the values assumed in ref.~\cite{sc76}.
The fit for the finite systems is quite
good, showing that rather simple considerations are sufficient to
describe the Kohn-Sham single particle energies.
Our fit parameters would give a band gap of 0.86 eV in infinitely
long polyene molecules.

The collective transition in the TDLDA has more complex behavior 
for large polyenes.  We find, unlike the carbon chains, that
the $\pi-\pi^*$ strength is partly split in the larger polyenes.
This is shown in Fig.~\ref{c8,18,28}, showing the low-energy TDLDA
strength function for N=8,18, and 28.  The systematics of
the energy of the strongest states is also shown in Fig.~4
as the crosses.  We have not found a theoretically motivated 
analytic parametrization for the $n$-dependence of these transitions.
The PPP empirical theory \cite{sc76}, when fitted to the smaller
polyenes, gives too high an energy for the largest ones, as is shown
by the dotted line in Fig.~4.  In Fig.~6
we show the comparison with experiment and with the linear response
calculations of ref. \cite{lu94,lu95}.  The two calculations agree
well with each other for the lighter systems, but they severely diverge
for the heavier molecules.  In principle, the small-amplitude TDLDA and 
the linear response are just different mathematical formulations of
the same theory, and they should give the same results for the same
Kohn-Sham energy functional.  Perhaps the differences are due to
different assumed bond lengths, as mentioned in the footnote. The experimental 
data for
polyenes in solution is shown by the crosses and squares.  There
is also gas phase data quoted in ref. \cite{lu95} which gives 10\%
higher energies.  From Fig.~6 one sees good agreement with the
lighter polyenes, but the data falling between the two theoretical
calculations for the heavier ones.   

The predicted transition strengths for the two calculations are also
significantly different.  Fig.~7 shows the oscillator strengths in
our calculation, compared to the nominal strength in the $\pi$ manifold.
The HOMO-LUMO transition, shown by the triangles, has roughly 
half the nominal strength.  This is further decreased in the TDLDA
by two effects.  One is the screening by the $\sigma$ electrons,
and the other is the loss of strength in the heavier polyenes to
other states.  In contrast, the calculation of \cite{lu95} gave
a collective transition with about 75\% of the nominal strength for
the heavier polyenes.  The stronger collectivity then gives a
larger upper shift in the energy from the HOMO-LUMO gap position. 

As a final example closely related to the polyenes, 
we examine retinal, the molecule of 
biological prominence as a retinyl chromophore.  Retinal is
more complicated than a polyene, having an aromatic termination
on one end and an carboxyl group on the other, but these play a minor
role in the collective longitudinal excitation. We consider
the  all-{\it trans} form, which is the final state of the
chromophore having absorbed a photon.  The experimental
data\cite{bi82} on {\it all-trans}-retinal shows broad, asymmetric 
peak with a maximum at 3.2 eV and a full width at half maximum of
about 0.6 eV. There is a previous {\it ab initio} study 
using the CASSCF method\cite{me97} which we can also compare to.
In our
calculation, the HOMO-LUMO gap is 2.0 eV, and this state is very strong
in the single-electron approximation.  The electron-electron
interaction in the TDLDA shifts the center of gravity of the strength up
to 3.3 eV, close to the experimentally observed position. In contrast,
the calculation of ref.~\cite{me97} obtained the strong transition
at much higher energy.  
However, as may be seen from Fig. \ref{retinal}, the
TDLDA strength is fragmented into three states.  The LDA predicts
the existence of a number of particle-hole states in the 3 eV
energy region, and the residual electron interaction is strong
enough to mix them.  The lowest predicted strong state, at
2.5 eV, is in a spectral region where the empirical absorption
strength is small.  This is similar to the problem we had with
the large polyenes.  In the retinal case, however, the situation is
complicated by the presence of an oxygen atom on the molecule.
As discussed in ref.~\cite{me97}, the lowest excitation in the molecule
has a predominant component with
a hole at the oxygen.  We have not checked that our mesh size
$\Delta x=0.3$~\AA~is small enough to describe oxygen reliably, and
that needs to be checked before making a final conclusion on the 
accuracy of the TDLDA for this system.  The transition strength associated with
the three strong transitions is $f=1.7$ in the TDLDA, which is close
to the prediction of ref.~\cite{me97}, but twice our extracted
experimental value.  That was obtained using the extinction
coefficient data of ref.~\cite{bi82}, correcting the solution data by
the Debye factor\cite{de29}.

\section{Simple aromatic molecules}
We discuss here the optical absorption spectra of benzene and
\C60.
\subsection{Benzene}
    Benzene provides our best example of the TDLDA comparisons to 
experiment. For this calculation, we take the empirical geometry of
benzene, $D_6$ symmetry with
CC bond length of 1.396 \AA~and CH bond length of  1.083 \AA.  The static
LDA calculation gives an energy spectrum with a 
gap $\Delta e $ of 5.04 eV.  This may be compared with
the empirical energy of the $^1B_{2u}$ excitation, which is a \pipi~
particle-hole state having a small residual interaction. That energy
is 4.9 eV\cite{hi91}, close to the LDA prediction.  
Other recent calculations of benzene excitations have been made using
the CASSFC and MRMP methods\cite{ha96}.  These authors find that
the $^1B_{2u}$ is also well reproduced by the CASSFC but not so well
by the MRMP.

Turning to the dynamic response,
the TDLDA transition strength 
is shown in Fig.~\ref{benzene}.
One sees the collective \pipi~state as a narrow peak at 6.9 eV.  
Thus the residual interaction pushes the \pipi~strength up 1.9 eV 
from its original position at $\Delta e$.
Above 9 eV there is a 
broad feature peaking around 18 eV.  This is associated with
the $\sigma-\sigma^*$
transitions; the width is 
due to the high level density of particle-hole states in the region of 
the resonance, and the fact that they are unbound.  However, our
implementation of TDLDA uses a finite sized box, which does not
give a satisfactory treatment of continuum effects. This spectrum
was calculated putting an absorptive boundary condition on the
wave function at the edge of the box.  Instead of the very 
fine structure predicted with a reflecting boundary condition, the
absorptive evidence produces (spurious) broader structures.

To compare with the empirical strength function, we consider 
data reported in two experiments.  In ref.~\cite{hi91}, the 
absolute strength was measured for excitation energies up to 
about 10 eV.  There is a quite sharp state at 6.9 eV in very good
agreement with our predicted \pipi~transition.  Comparing with
the other theoretical methods, the strong state is well reproduced by
the the MRMP but not the CASSCF\cite{ha96}.  
The
empirical strength integrating the peak region from 6.5 eV up to the 
minimum at 8.3 eV is $f=0.9$ eV.  This agrees well with our theoretical 
strength for
the collective transition, $f=1.1$.  Here both the CASSCF and MRMP
give even better agreement with experiment.  We mention that none
of the theories describe the finer details of the absorption
spectrum around the 6.9 eV peak, which has an asymmetric tail 
and a small peak on the low-energy side.  Since
theory predicts a single sharp state, it is clear that other degrees
of freedom, probably the vibrational, are responsible for the detailed
structure of the peak. 
The experimental strength function
also show very fine peaks in the neighborhood of the ionization energy.
These are Rydberg states which are beyond the reach of simple energy
densities functions, which lack the Coulomb tail in the mean field.
Another experiment\cite{ko72} reported the absorption strength going up 
to much higher
energies, but without an absolute normalization. In Fig.~\ref{benzene}
we show this data normalized to our calculation.
We see that broad $\sigma -\sigma^*$ feature is reproduced in position
and width very well. However as mentioner, fine details of the spectrum 
are not properly reproduced.

\section{\C60}
Our last example is the molecule \C60.  This molecule has
been well studied in the LDA; the predicted HOMO-LUMO gap is
1.65 eV, compared with the measured gap of $\sim 1.6$~eV.
The strength function computed with the TDLDA is shown in 
Fig.~\ref{c60-1}.  We see a set of strong transitions in the
region 4-7 eV excitation, and a much larger and broad set in the
region of 15-30 eV.  The lower transitions have a  \pipi
character.  Experimentally, the absorption strength below 6 eV
has been measured in absolute terms and the data is reviewed in
ref. \cite{sm96}.  Comparison of theory and experiment is
made in Table  I.  There is a good correspondence between the TDLDA
transition at 3.4 eV and the observed state at 3.8 eV. For the higher transitions, theory predicts more 
strong states than the number of peaks observed, so there is
no one-to-one correspondence.  However, the overall strength distribution
comes out fairly well.  This may be seen in Fig.~\ref{c60-2} comparing
the integrated strength distribution up to 6.2 eV.  Of course the
TDLDA strength is sharply peaked without the line-broadening
processes that would be computable in a more complete theory.
The total strength in the region comes out about right in TDLDA,
but shifts of the order of 0.5 eV would be required to fit the data.

The strong absorption between 15 and 30 eV is due to
$\sigma-\sigma^*$ transitions.  Of course, the sharp structures
predicted in the TDLDA would be smeared out due to coupling to
the continuum and to vibrations.  The only data
in this region is the photoelectron cross sections\cite{he92}.
It shows a peak at 20 eV having a full width at half
maximum of about 10 eV, consistent with the TDLDA results.

\section{Conclusion}
The TDLDA is an attractive theory for collective excitations 
in large molecules because of its numerical scaling properties,
and its automatic compliance with sum rules.
We have found it to be rather good for conjugated carbon molecules, 
giving a reasonable account of 
the strong transitions in both the \pipi~ and \sigsig~manifolds
of states.  

In the \pipi~manifold, the energetics of the underlying single-particle
states are fairly well described by the H\"uckel Hamiltonian.  The
results are summarized in Fig.~\ref{hueckel}, showing the deduced 
hopping matrix elements as a function of the CC bond length.  
Fitting the
conjugated molecules to a power law dependence on bond length, the
extracted matrix elements vary as $t\sim r^{-2.7}$.  This is 
a stronger dependence than the commonly assumed form, $t\sim r^{-2}$.
The matrix element for the carbon chain molecules is quite different, perhaps
because of the absence of hydrogen atoms.
  
As is well known, the electron-electron interaction strongly perturbs the
absorption strength function.  In the region of the \pipi~transitions,
the strength is shifted upward by an amount of the order of 1-2 eV
for the systems studied.  In the case of carbon chains, the transition
energy can be described fairly well analytically.  The energies agree
well with the empirical values for the smaller molecules, but the larger
ones are predicted to have more fragmentation of strength in the
TDLDA than is observed.  The amount of strength in the \pipi~transition
in principle depends on the coupling to the other electrons.  In
small or extended molecules, the  coupling and resulting 
screening is rather mild, less than a factor of two.  On the
other hand, in \C60 the \pipi~transition strength is more strongly 
screened. 
   The absorption spectra typically have a gap above the the strong
\pipi~transition, followed by a dense spectrum of \sigsig~transitions
above 10 eV.  This spectrum peaks in the range of 15-20 eV and is
quite broad.  This feature is present in the empirical absorption,
but quantitative experimental data is only available for one case,
benzene.  There we found good agreement on the overall shape and
size of the \sigsig~absorption peak.
   In view of the evident success of the TDLDA theory, it would be
interesting to consider more ambitious applications.  One direction
is to try other, more difficult, elements.  In particular, it may
be possible to treat $d$-shell elements with mesh sizes of the
order of $\Delta x\approx 0.2$~\AA, which would allow small
molecules and clusters to be calculated. Another direction would
be to study the effect of perturbations on the spectra.  Possible
perturbations include:  nearby charges or nearby molecules; thermal
distributions of ionic geometries; electron-vibration coupling.

This work is supported in part by 
the Department of Energy  under Grant DE-FG-06-90ER40561, and by the
Grant-in-Aid for Scientific Research from the Ministry of Education, 
Science and Culture (Japan), No. 09740236. Numerical calculations were
performed on the FACOM VPP-500 supercomputer in the institute for
Solid State Physics, University of Tokyo, and on the NEC sx4 supercomputer
in the research center for nuclear physics (RCNP), Osaka University.

\begin{table}
\caption{Oscillator strengths $f$ for the near-ultraviolet transitions in
\C60}
\begin{tabular}{cccc}
    TDLDA     & TDLDA         &    experiment         &  experiment\\
E (eV)  &  $f$ & E (eV) & $f$ \\
\tableline
3.4 &   0.3   & 3.8 & 0.5 \\
4.3 &   0.9   &   &  \\
5.3 &  2.5    &  4.8 & 2.1-2.5 \\
6.0 &  2.5  &  5.8 & 5-6 \\
6.6 & 1.1   &      &     \\
\end{tabular}
\end{table}
\newpage
\section*{Figure Captions}
{Fig.~1 Computation times for the TDLDA method.  Shown are the
times in minutes for computation of polyenes on the NEC sx4
vector process computer, with an effective speed of 0.4 Gflop,
as a function of the number of of electrons
in the molecule.  Crosses show the time to construct the ground
state, taking 1500 iteration steps. Open triangles show the duration 
of the time-evolution phase of the calculation, taking 10000 iteration
steps.A quadratic functional dependence is displayed with the lines.}

{Fig.~2 C$_8$H$_{10}$ orbitals with $\pi$ character.  The left-hand
side shows the results of the static LDA calculation.
The spectrum of the H\"uckel Hamiltonian with matrix elements given
by eq.~(10) is shown on the right.  The energies of the H\"uckel
spectrum have been shifted to match the
position of the HOMO state.
}

Fig.~3 Strength function of C$_{8}$H$_{10}$ in the energy
region 0-30 eV, calculated in the
TDLDA.

Fig.~4 Results of the present TDLDA calculations for polyenes,
C$_n$H$_{n+2}$.
The HOMO-LUMO gap is shown by the triangles, and the fit to
the generalized H\"uckel Hamiltonian with matrix elements eq.~(10) 
by the dashed line.  For comparison, the $n^{-1}$ 
dependence similar to eq.~(7)  is
shown by a dot-dashed line. The energies of the strongest transition
in TDLDA are shown by crosses.   
The strong transition in the PPP empirical theory is shown by
the dotted line.

Fig.~5 TDLDA strength function for low-energy transitions in the polyenes
with $n=8,18$ and 28.

Fig.~6~Energy of the collective $\pi-\pi^*$ transition in
polyenes, comparing theory with
experiment.  The solid line connects the energies found from the
TDLDA, and the dashed line connects the results of \protect\cite{lu95}.
Crosses and squares are data in solution from ref. \protect\cite{gr81} 
and \protect\cite{so61}, 
and triangles are gas data as quoted by ref. \protect\cite{lu95}.

Fig.~7 Strengths of the collective $\pi-\pi^*$ transitions in polyenes:
single-electron HOMO-LUMO transition, triangles; collective TDLDA 
transition, crosses. The total strength associated with the $\pi$
manifold is shown by the dashed line.

Fig.~8 Strength function for retinal; TDLDA, solid; experiment
\protect\cite{bi82}, dashed.

Fig.~9 Benzene orbital energies in the static LDA.  On the
far right are the $\pi$ energies in the H\"uckel model, with $\beta=2.52$~eV,
shifted to match the HOMO energy.

Fig.~10 Optical absorption of the benzene molecule, in units of
eV$^{-1}$. a) experimental, from
ref. \protect\cite{ko72}; b) TDLDA.  An absorptive potential has been
put at the boundaries of the grid to mimic the continuum. The grid 
has the shape of a sphere of radius 7 \AA~with $\Delta x=0.3$ \AA~and
$\Delta t= 0.001$ eV$^{-1}$.  The number of spatial mesh points is
about 50,000. The
figure shows the Fourier tranform of
real-time response over a time $T=30$ eV$^{-1}$.

Fig.~11. Strength for \C60 calculated in the TDLDA.

Fig.~12. Integrated strength below 6.2 eV in \C60, TDLDA, dashed line;
experiment \protect\cite{sm96}, solid line.

Fig.~13. Effective H\"uckel matrix elements in the
molecules studied.  They are, in ordering of increasing bond length: carbon
chains,polyene single bond, benzene, and polyene double bond.
\newpage

\begin{figure}
  \begin{center}
    \leavevmode
    \parbox{0.9\textwidth}
           {\psfig{file=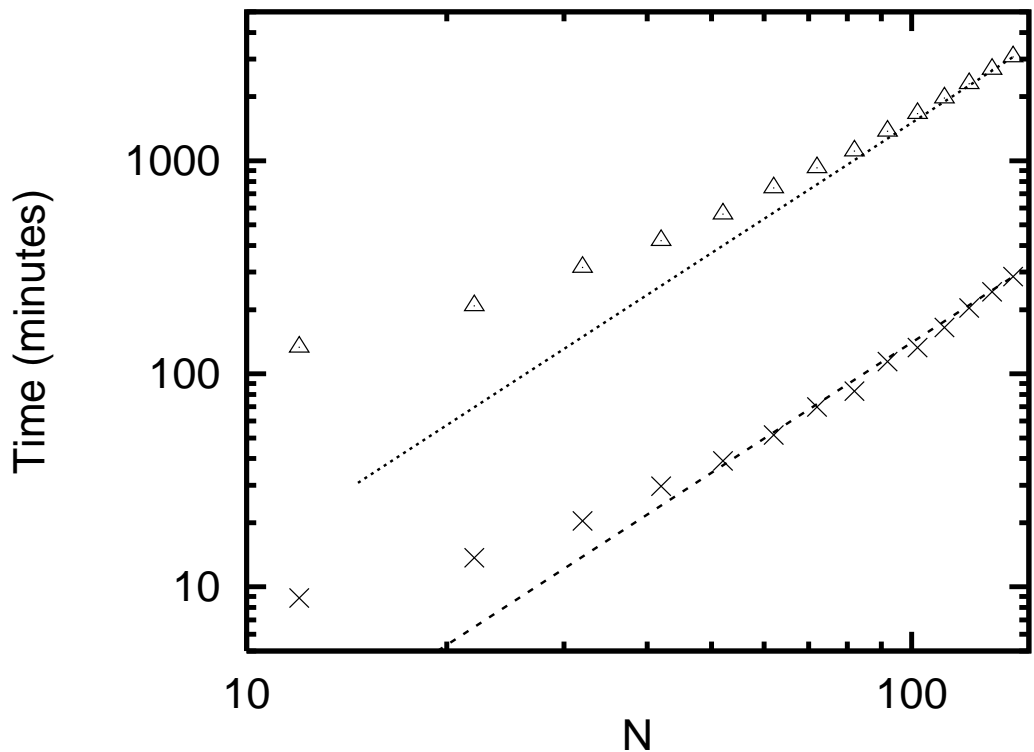,width=0.9\textwidth}}
  \end{center}
\protect\caption{Computation times for the TDLDA method.  Shown are the
times in minutes for computation of polyenes on the NEC sx4
vector process computer, with an effective speed of 0.4 Gflop,
as a function of the number of of electrons
in the molecule.  Crosses show the time to construct the ground
state, taking 1500 iteration steps. Open triangles show the duration 
of the time-evolution phase of the calculation, taking 10000 iteration
steps.A quadratic functional dependence is displayed with the lines.}
\label{timing}
\end{figure}

\begin{figure}
  \begin{center}
    \leavevmode
    \parbox{0.9\textwidth}
           {\psfig{file=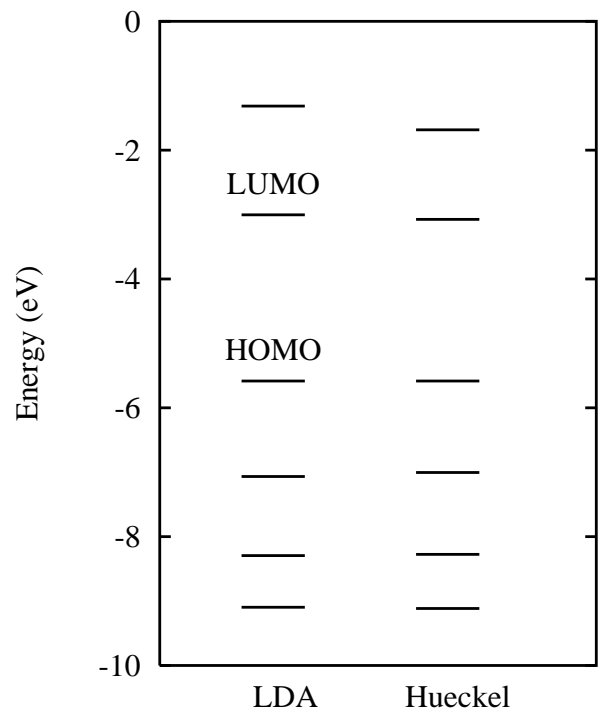,width=0.5\textwidth}}
  \end{center}
\caption{C$_8$H$_{10}$ orbitals with $\pi$ character.  The left-hand
side shows the results of the static LDA calculation.
The spectrum of the H\"uckel Hamiltonian with matrix elements given
by eq.~(10) is shown on the right.  The energies of the H\"uckel
spectrum have been shifted to match the
position of the HOMO state.
}
\label{polyene-energies}
\end{figure}

\begin{figure}
  \begin{center}
    \leavevmode
    \parbox{0.9\textwidth}
           {\psfig{file=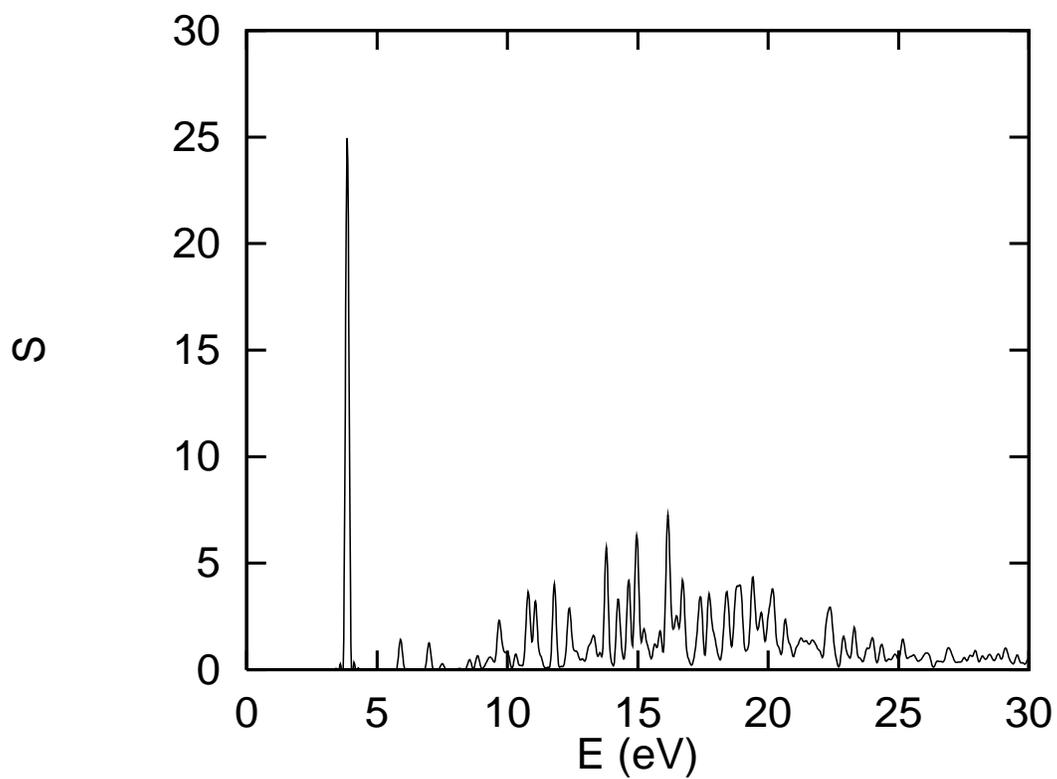,width=0.9\textwidth}}
  \end{center}
\caption{Strength function of C$_{8}$H$_{10}$ in the energy
region 0-30 eV, calculated in the
TDLDA.}
\label{c8h10}
\end{figure}
\newpage
\begin{figure}
  \begin{center}
    \leavevmode
    \parbox{0.9\textwidth}
           {\psfig{file=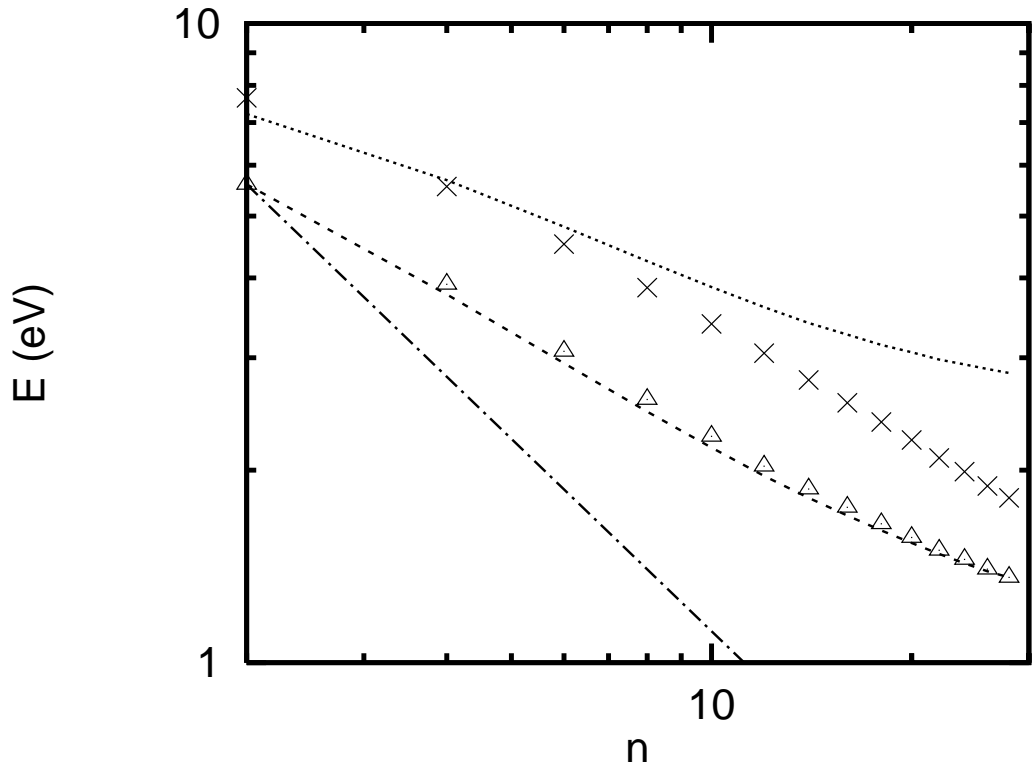,width=0.9\textwidth}}
  \end{center}
\caption{Results of the present TDLDA calculations for polyenes,
C$_n$H$_{n+2}$.
The HOMO-LUMO gap is shown by the triangles, and the fit to
the generalized H\"uckel Hamiltonian with matrix elements eq.~(10) 
by the dashed line.  For comparison, the $n^{-1}$ 
dependence similar to eq.~(7)  is
shown by a dot-dashed line. The energies of the strongest transition
in TDLDA are shown by crosses.   
The strong transition in the PPP empirical theory is shown by
the dotted line.
}
\label{polyene1}
\end{figure}

\begin{figure}
  \begin{center}
    \leavevmode
    \parbox{0.9\textwidth}
           {\psfig{file=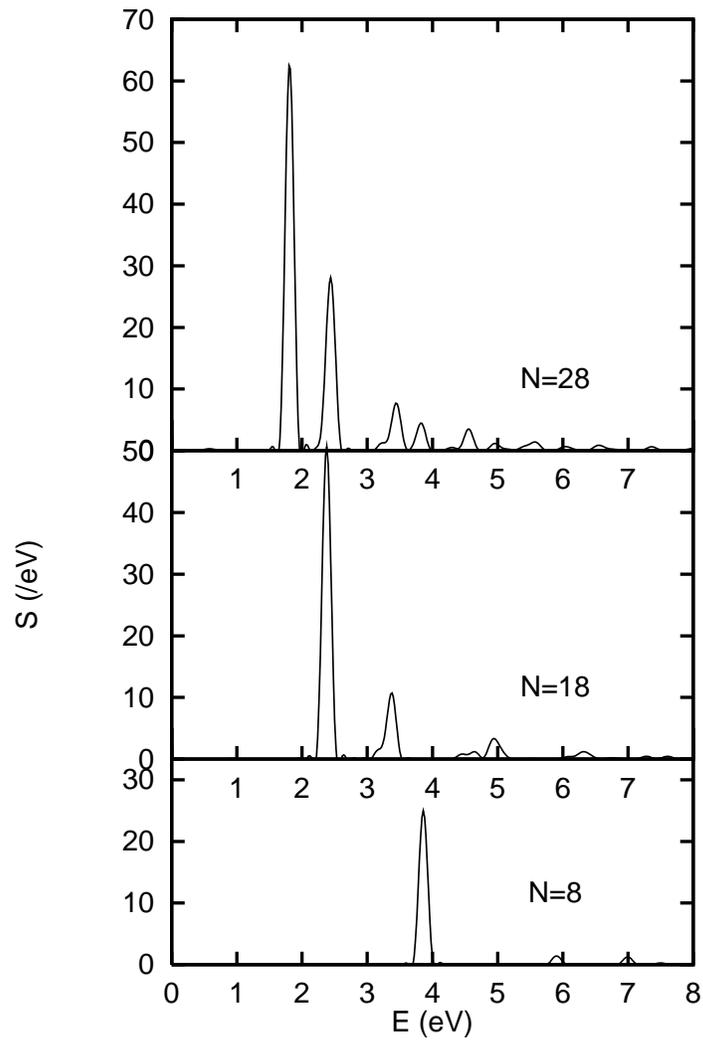,width=0.6\textwidth}}
  \end{center}
\caption{TDLDA strength function for low-energy transitions in the polyenes
with $n=8,18$ and 28.}
\label{c8,18,28}
\end{figure}

\begin{figure}
  \begin{center}
    \leavevmode
    \parbox{0.9\textwidth}
           {\psfig{file=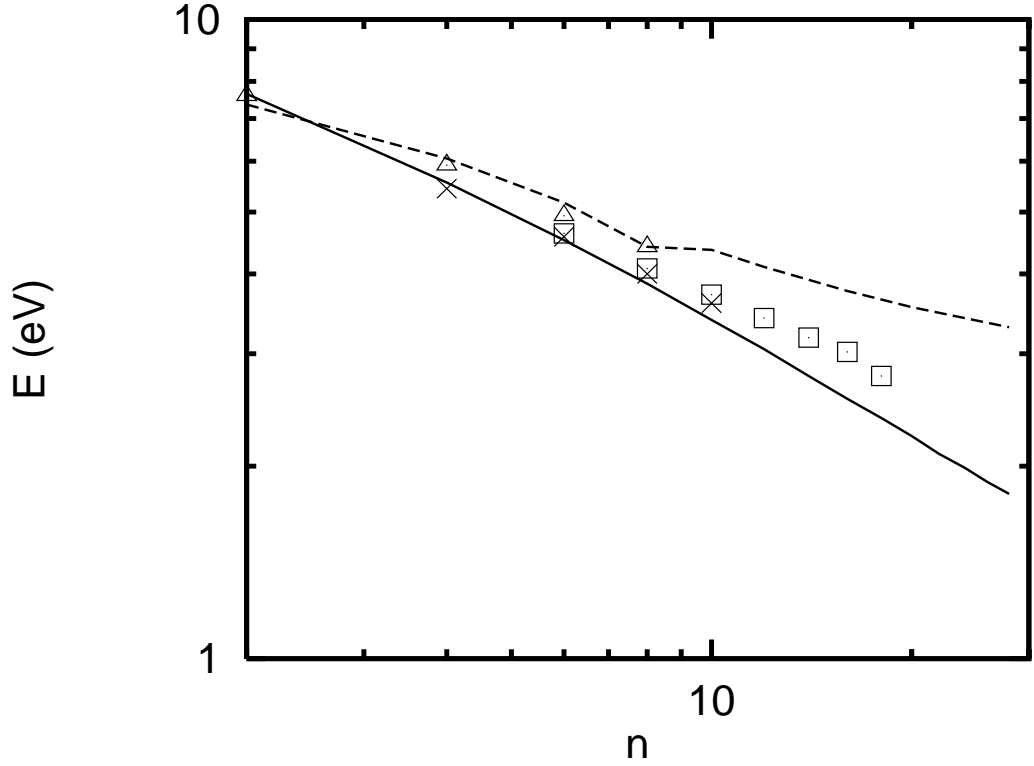,width=0.9\textwidth}}
  \end{center}
\protect\caption{Energy of the collective $\pi-\pi^*$ transition in
polyenes, comparing theory with
experiment.  The solid line connects the energies found from the
TDLDA, and the dashed line connects the results of \protect\cite{lu95}.
Crosses and squares are data in solution from ref. \protect\cite{gr81} 
and \protect\cite{so61}, 
and triangles are gas data as quoted by ref. \protect\cite{lu95}.
}
\label{compare}
\end{figure}

\begin{figure}
  \begin{center}
    \leavevmode
    \parbox{0.9\textwidth}
           {\psfig{file=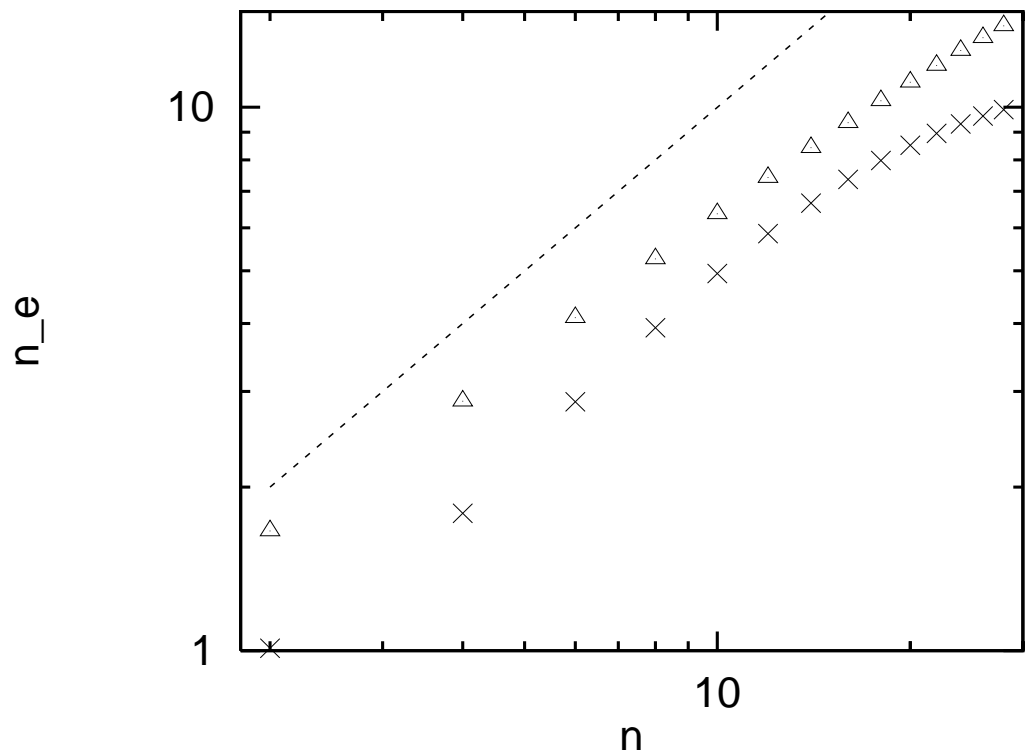,width=0.9\textwidth}}
  \end{center}
\caption{Strengths of the collective $\pi-\pi^*$ transitions in polyenes:
single-electron HOMO-LUMO transition, triangles; collective TDLDA 
transition, crosses. The total strength associated with the $\pi$
manifold is shown by the dashed line.
}
\label{transition strengths}
\end{figure}

\begin{figure}
  \begin{center}
    \leavevmode
    \parbox{0.9\textwidth}
           {\psfig{file=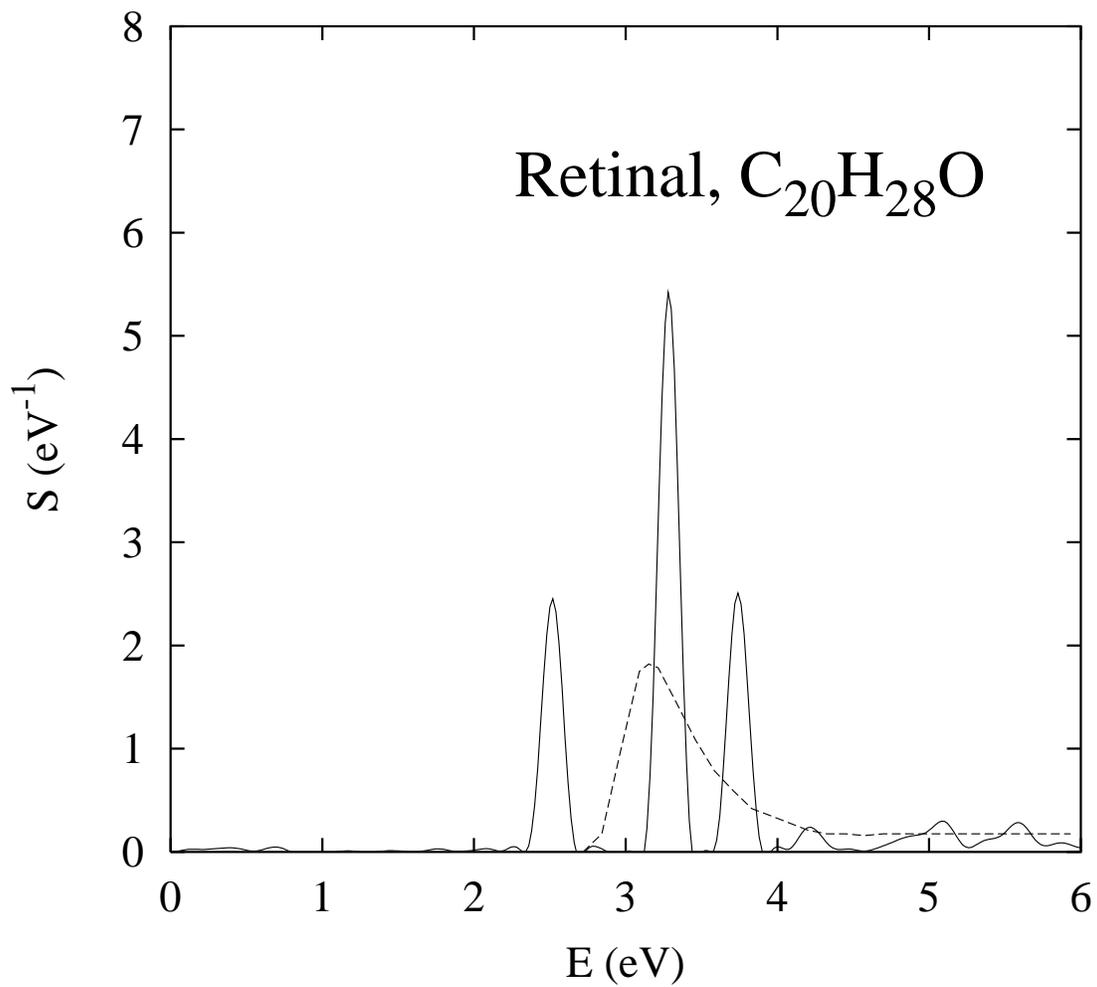,width=0.9\textwidth}}
  \end{center}
\protect\caption{Strength function for retinal; TDLDA, solid; experiment
\protect\cite{bi82}, dashed.}
\label{retinal}
\end{figure}

\begin{figure}
  \begin{center}
    \leavevmode
    \parbox{0.9\textwidth}
           {\psfig{file=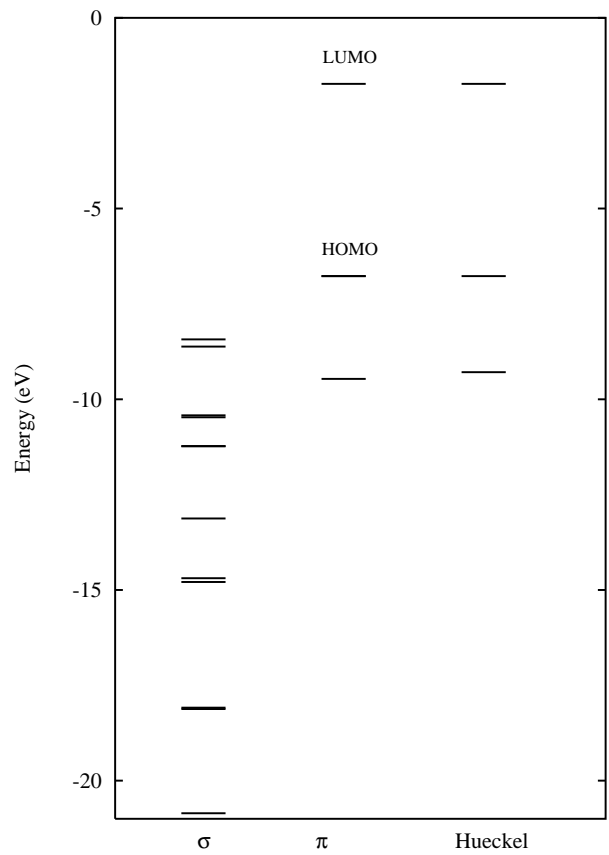,width=0.5\textwidth}}
  \end{center}
\protect\caption{Benzene orbital energies in the static LDA.  On the
far right are the $\pi$ energies in the H\"uckel model, with $\beta=2.52$~eV,
shifted to match the HOMO energy.
}
\label{benzene-hueckel}
\end{figure}

\begin{figure}
  \begin{center}
    \leavevmode
    \parbox{0.9\textwidth}
           {\psfig{file=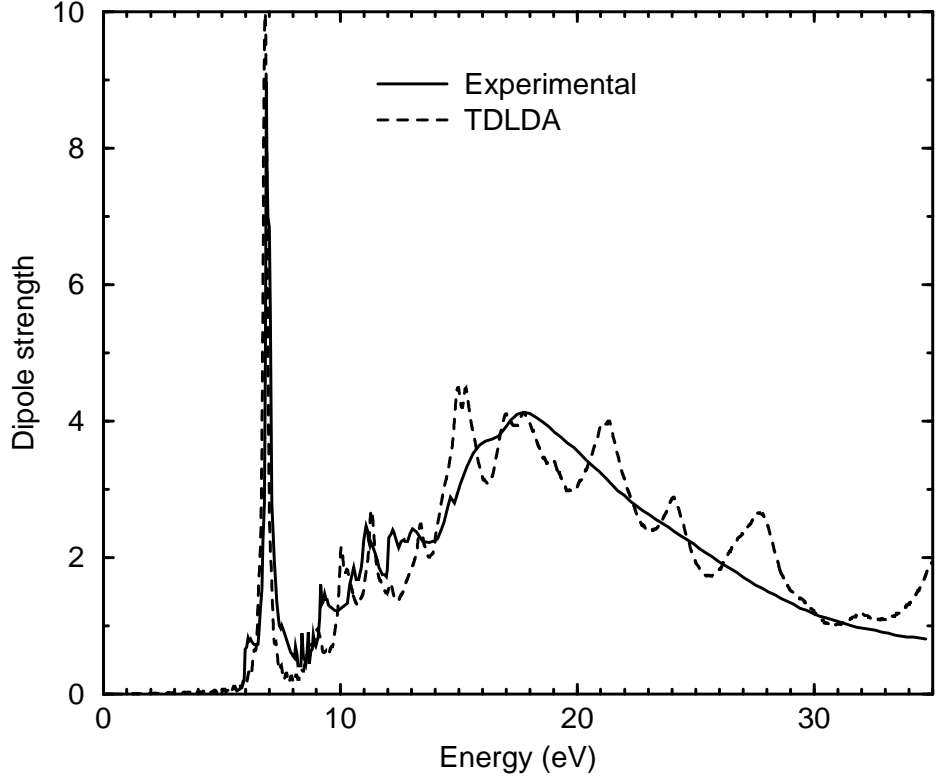,width=0.9\textwidth}}
  \end{center}
\caption{Optical absorption of the benzene molecule, in units of
eV$^{-1}$. a) experimental, from
ref. \protect\cite{ko72}; b) TDLDA.  An absorptive potential has been
put at the boundaries of the grid to mimic the continuum. The grid 
has the shape of a sphere of radius 7 \AA~with $\Delta x=0.3$ \AA~and
$\Delta t= 0.001$ eV$^{-1}$.  The number of spatial mesh points is
about 50,000. The
figure shows the Fourier tranform of
real-time response over a time $T=30$ eV$^{-1}$.}
\label{benzene}
\end{figure}

\begin{figure}
  \begin{center}
    \leavevmode
    \parbox{0.9\textwidth}
           {\psfig{file=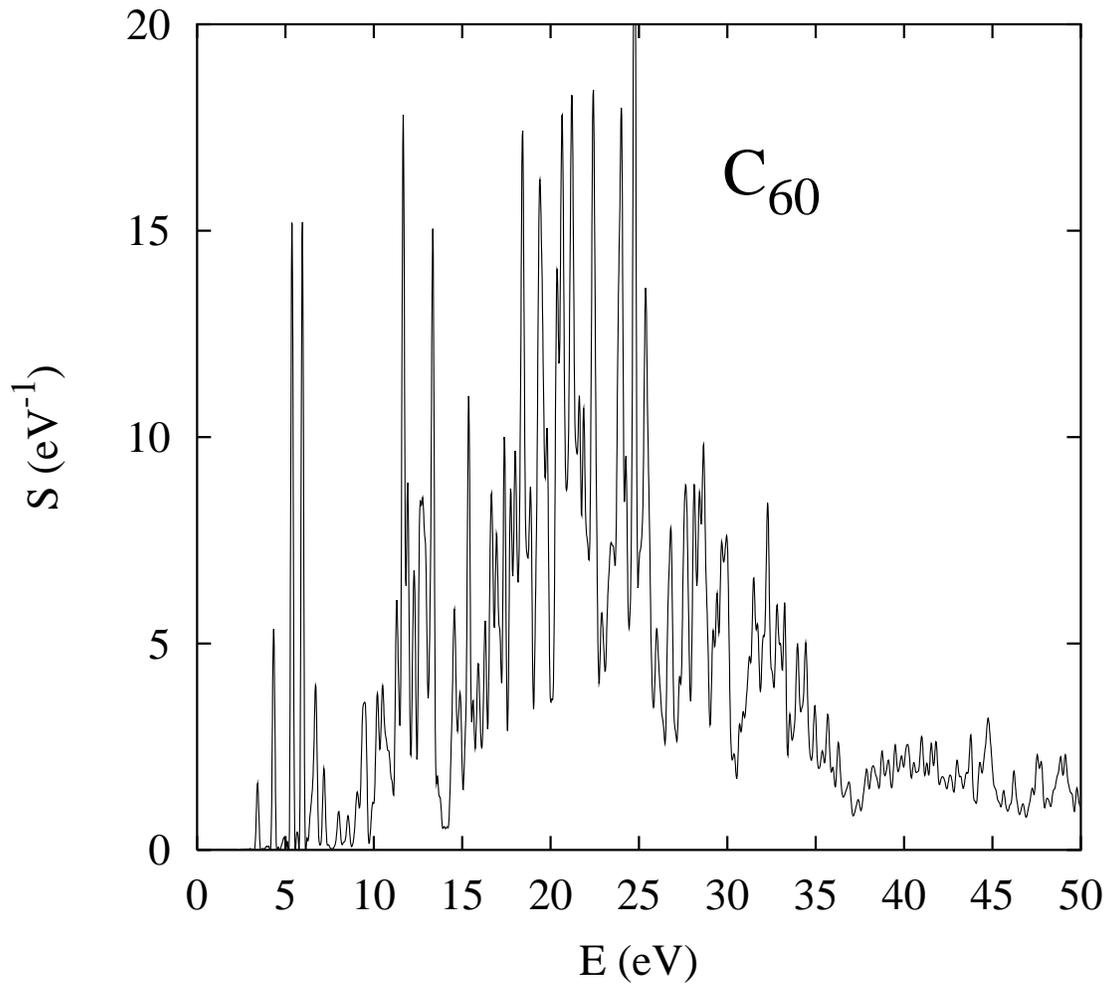,width=0.9\textwidth}}
  \end{center}
\caption{Strength for \C60 calculated in the TDLDA.}
\label{c60-1}
\end{figure}

\begin{figure}
  \begin{center}
    \leavevmode
    \parbox{0.9\textwidth}
           {\psfig{file=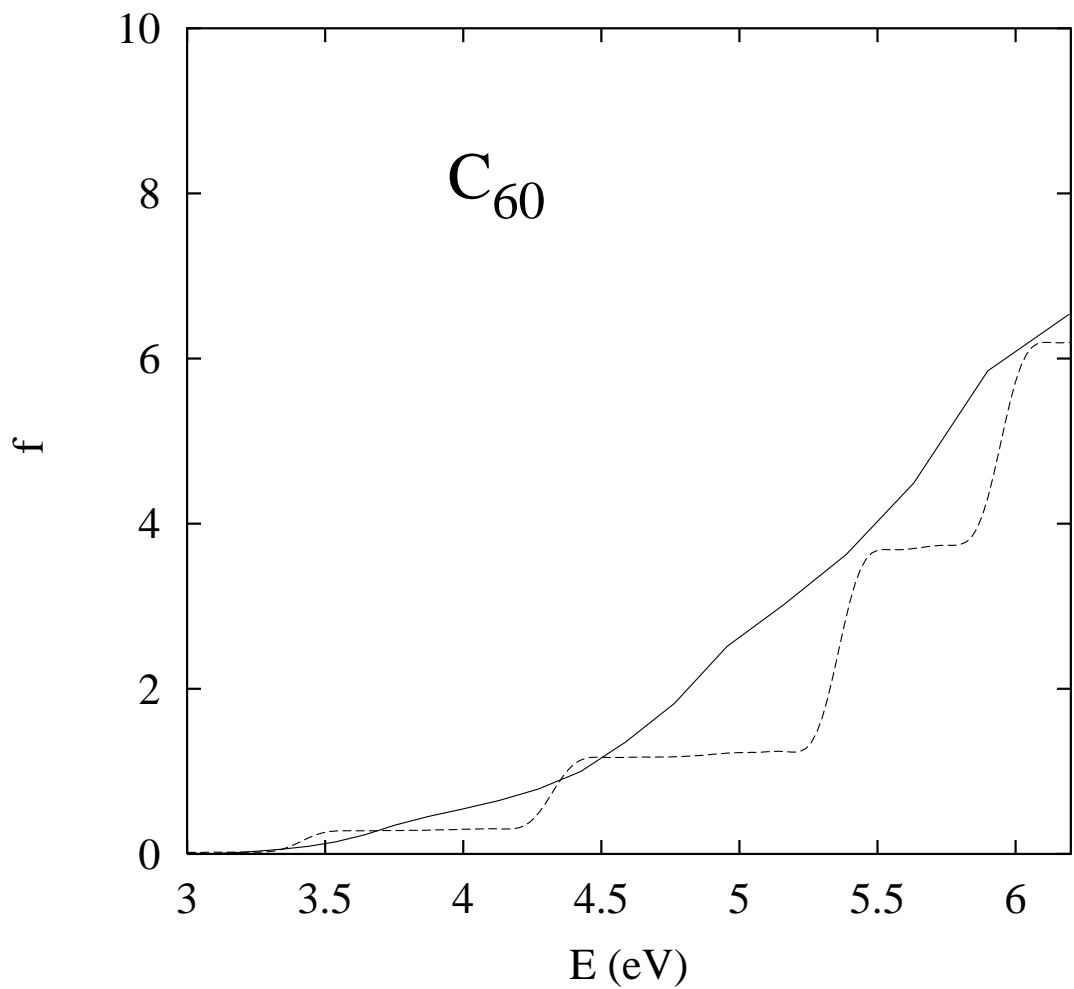,width=0.9\textwidth}}
  \end{center}
\caption{Integrated strength below 6.2 eV in \C60, TDLDA, dashed line;
experiment \protect\cite{sm96}, solid line.} 
\label{c60-2}
\end{figure}

\begin{figure}
  \begin{center}
    \leavevmode
    \parbox{0.9\textwidth}
           {\psfig{file=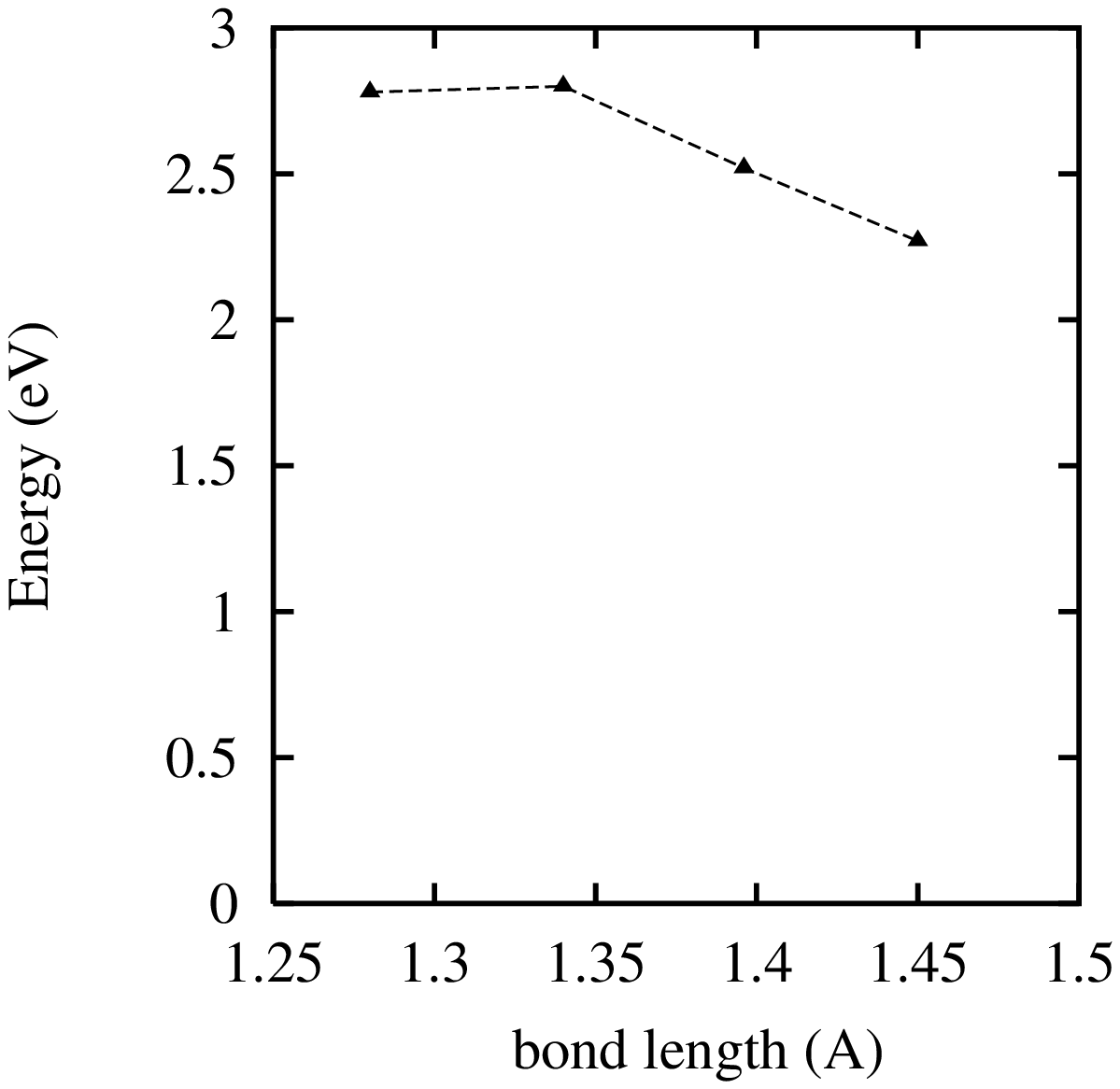,width=0.5\textwidth}}
  \end{center} \caption{Effective H\"uckel matrix elements in the
molecules studied.  They are, in ordering of increasing bond length: carbon
chains,polyene single bond, benzene, and polyene double bond.
} 
\label{hueckel} 
\end{figure}


\begin{references}
\bibitem{jo89} R.O. Jones, O. Gunnarsson, Rev. Mod. Phys. 61 689 (1989).
\bibitem{ya94} C. Yannouleas, et al., J. Phys. B27 L642 (1994).
\bibitem{ko93} H. Koch, et al., Chem. Phys. 172 13 (1993).
\bibitem{lu94} Y. Luo, et al., J. Phys. Chem. 98 7782 (1994).
\bibitem{lu95} Y. Luo, et al., Phys. Rev. B51 14949 (1995).
\bibitem{fe92} M. Feyereisen, et al., J. Chem. Phys. 96 2978 (1992).
\bibitem{ya96} K. Yabana and G.F. Bertsch, Phys. Rev. B54 4484 (1996).
\bibitem{ya97} K. Yabana and G.F. Bertsch, Z. Phys. D42 219 (1997).
\bibitem{ja96} C. Jamorski, et al, J. Chem. Phys. 104 5134 (1996).
\bibitem{bl95} X. Blase, et al., Phys. Rev. B52 R2225 (1995).
\bibitem{ru96} A. Rubio, et al., Phys. Rev. Lett. 77 247 (1996).
\bibitem{za80} A. Zangwill and P. Soven, Phys. Rev. A21 (1980) 1561.
\bibitem{di30} P.A.M. Dirac, Proc. Cambridge Phil. Soc. 26 376 (1930).
\bibitem{pe96} M. Petersilka, U. Grossmann, and E. Gross, Phys. Rev.
Lett. 76 1212 (1996).
\bibitem{de92} F.V. DeBlasio, et al., Phys. Rev. Lett. 68 1663 (1992).
\bibitem{vr97} D. Vretanar, et al., Nucl. Phys. A621 853 (1997).
\bibitem{ya93} C. Yannouleas, et al., Phys. Rev. B47 9849 (1993); F.
Alasia, et al., J. Phys. B27 L643 (1994).
\bibitem{ch94} J. Chelikowsky, N. Troullier, K. Wu, and Y. Saad, Phys. Rev.
B50 11355 (1994).
\bibitem{fl78} H. Flocard, S. Koonin, and M. Weiss, Phys. Rev. C17 1682
(1978).
\bibitem{ce80} D. Ceperley and B. Alder, Phys. Rev. Lett. 45 566 (1980).
\bibitem{pe81} J. Perdew and A. Zunger, Phys. Rev. B23 5048 (1981).
\bibitem{tr91} N. Troullier, J.L. Martins, Phys. Rev. B43 1993 (1991).
\bibitem{al97} P. Alippi, P. La Rocca, and G. Bachelet, Phys. Rev.
B55 13855 (1997).
\bibitem{ya98} K. Yabana and G.F. Bertsch, to be published in Phys. Rev. A;
Los Alamos preprint phys/9802017.
\bibitem{kl82} L. Kleinman and D. Bylander, Phys. Rev. Lett. 1425 (1982).
\bibitem{sa96} Y. Saad, et al., BIT 36 563 (1996).
\bibitem{mi59}
Y. Mizuno and T. Izuyama, Prog. Theo. Phys. 21 593 (1959).
\bibitem{pa88}
G. Paacchioni and J. Koutecky, J. Chem. Phys. 88 1066 (1988).
\bibitem{ko95}
M. Kolbuszewski, J. Chem. Phys. 102 3679 (1995).
\bibitem{sc76}
K. Schulten, I. Ohmire, and M. Karplus, J. Chem. Phys. 64 4422 (1976).
\bibitem{so61} F. Sondheimer, et al., J. Am. Chem. Soc. 83 1675 (1961).
\bibitem{gr81} M.F. Granville, et al., J. Chem. Phys. 75 3765 (1981).
\bibitem{bi82} R.R. Birge, et al., J. Am. Chem. Soc. 104 1196 (1982).
\bibitem{me97} M. Merchan and R. Gonzales-Luque, J. Chem. Phys. 106 1112
(1997).
\bibitem{de29}
P. Debye, Polar Molecules (Chemical Catalog, New York, 1929).
\bibitem{hi91} A. Hiraya and K. Shobatake, J. Chem. Phys. 94 (1991) 7700. 
\bibitem{ha96} T. Hashimoto, H. Nakano, and K. Hirano, J. Chem. Phys. 104
6244 (1996).
\bibitem{ko72} E.E. Koch and A. Otto, Chem. Phys. Lett. 12 (1972) 476.   
\bibitem{sm96} A.L. Smith, J. Phys. B29 4975 (1996).
\bibitem{he92}  I.V. Hertel et al., Phys. Rev. Lett. 68 784 (1992).
\end{references}
\end{document}